\pdfoutput=1
\documentclass[a4paper, 12pt]{article}
\usepackage[dvipsnames]{xcolor}

\makeatletter  
\gdef\@fpheader{}  
\makeatother 

\usepackage[T1]{fontenc} 
\usepackage{tangocolors}
\usepackage{tikz}
\usepackage{tikz-3dplot}
\usetikzlibrary{arrows, decorations,backgrounds, patterns, positioning,automata}
\usetikzlibrary{decorations.pathreplacing ,decorations.markings}
\usepackage{amssymb}
\usetikzlibrary{decorations.pathmorphing,backgrounds,shapes,arrows,shadows}
\usetikzlibrary{patterns.meta}
\usetikzlibrary{patterns,decorations.pathmorphing}
\usetikzlibrary{}

\tikzset{
    snake it/.style={decorate, decoration=snake}
}
\usepackage{pgfplots}
\pgfplotsset{compat=1.11}
\usepgfplotslibrary{fillbetween}
\usetikzlibrary{intersections}
\pgfdeclarelayer{bg}
\pgfsetlayers{bg,main}
\tikzset{zigzag/.style={decorate,decoration=zigzag}}
\tikzset{snake it/.style={decorate, decoration=snake}}
\makeatletter
\def\@hex@@Hex#1%
 {\if a#1A\else \if b#1B\else \if c#1C\else \if d#1D\else
  \if e#1E\else \if f#1F\else #1\fi\fi\fi\fi\fi\fi \@hex@Hex}
\makeatother

\usepackage[all]{xy}
\usepackage[percent]{overpic}
\usepackage{slashed}
\usepackage{wrapfig}
\usepackage{tabu}
\usepackage{diagbox}
\usepackage{mathrsfs,amsmath,amssymb,amsthm,amsfonts,tikz,graphicx,accents,hyperref, color}
\usepackage{dsfont,epiolmec, latexsym, stmaryrd, comment}
\usepackage{slashed,ccaption}
\usepackage{mathrsfs, calligra}
\usepackage{leftidx}
\usepackage{import}
\usepackage{multirow}
\usepackage{amsfonts}
\usepackage{pifont}
\usepackage{tabularx}
\usepackage{cancel}
\usepackage[utf8]{inputenc}
\usetikzlibrary{intersections,calc}
\usepackage{ifthen}
\usepackage{amsmath}
\usepackage{physics}
\usepackage{cancel}
\usepackage{caption}
\usepackage{subcaption}

\usetikzlibrary{patterns.meta}
\usepackage{array}
\usepackage{bm}
\usepackage[T1]{fontenc}
\usepackage{lmodern}
\let\sho\th

\usepackage[most]{tcolorbox}

\tcbset{highlight math style={left=02mm,right=02mm,top=-1mm,bottom=-1mm}} 
\usepackage{empheq}

\hypersetup{ linktoc=all,
    colorlinks, linkcolor={palatinateblue},
    citecolor={brightpink}, urlcolor={amaranth}
}

\graphicspath{{Images/}}

\def\sideremark#1{\ifvmode\leavevmode\fi\vadjust{\vbox to0pt{\vss
 \hbox to 0pt{\hskip\hsize\hskip1em
 \vbox{\hsize2cm\tiny\raggedright\pretolerance10000
 \noindent #1\hfill}\hss}\vbox to8pt{\vfil}\vss}}}%
\DeclareSymbolFont{extraup}{U}{zavm}{m}{n}
\DeclareMathSymbol{\varheart}{\mathalpha}{extraup}{86}
\DeclareMathSymbol{\vardiamond}{\mathalpha}{extraup}{87}
\makeatletter
\renewcommand*{\@fnsymbol}[1]{\ensuremath{\ifcase#1\or \clubsuit \or \vardiamond \or \varheart\or
    \spadesuit\or \mathparagraph\or \|\or **\or \dagger\dagger
    \or \ddagger\ddagger \else\@ctrerr\fi}}
\makeatother

\definecolor{rosy}{RGB}{230,235,252}
\definecolor{myframetitle}{RGB}{90,89,170}
\definecolor{myblocktitle}{RGB}{140,185,249}
\definecolor{mytitle}{RGB}{10,80,26}

\definecolor{darkgreen}{RGB}{27,130,45}
\definecolor{darkblue}{rgb}{0,0,0.3}
\definecolor{darkred}{rgb}{0.7,0,0}

\definecolor{lightred}{rgb}{0.1,.05,0.05}
\definecolor{lightblue}{rgb}{0,.1,0.05}

\definecolor{light gray}{RGB}{220,220,220}
\definecolor{dark purple}{RGB}{108,0,217}
\definecolor{pink}{RGB}{190,20,100}
\definecolor{orang}{RGB}{193,63,0}
\definecolor{green}{RGB}{11,98,17}
\definecolor{darkpink}{RGB}{153,0,76}
\definecolor{bluegreen}{RGB}{0,102,102}
\definecolor{greenlagan}{RGB}{0,102,0}
\definecolor{redgreen}{RGB}{102,102,0}
\definecolor{Redgreen}{RGB}{153,76,0}
\definecolor{vividviolet}{rgb}{0.62, 0.0, 1.0}
\definecolor{amaranth}{rgb}{0.9, 0.17, 0.31}
\definecolor{palatinateblue}{rgb}{0.15, 0.23, 0.89}
\definecolor{brightpink}{rgb}{1.0, 0.0, 0.5}
\definecolor{cornflowerblue}{rgb}{0.39, 0.58, 0.93}
\definecolor{deepcarminepink}{rgb}{0.94, 0.19, 0.22}
\definecolor{radicalred}{rgb}{1.0, 0.21, 0.37}

\DeclareFontFamily{OT1}{rsfs}{}

\DeclareFontShape{OT1}{rsfs}{m}{n}{ <-7> rsfs5 <7-10> rsfs7 <10->rsfs10}{}

\DeclareMathAlphabet{\mycal}{OT1}{rsfs}{m}{n}

\newcommand{\cHp}{\mathcal{H}_{\text{p}}}
\newcommand{\cHt}{\mathcal{H}_{\text{tot}}}
\newcommand{\cHc}{\mathcal{H}_{\text{c}}}
\newcommand{\bcO}{\boldsymbol{\mathcal{O}}}
\newcommand{\bcC}{\boldsymbol{\mathcal{C}}}

\newcommand{\be}{\begin{equation}}
\newcommand{\ee}{\end{equation}}
\newcommand{\bea}{\begin{eqnarray}}
\newcommand{\eea}{\end{eqnarray}}
\newcommand\tcr{\textcolor{red}}

\begin{document}

\begin{titlepage}
\begin{center}
 {\textbf{\Large{Revisiting Quantization of  Gauge Field Theories: \\ \vskip 2mm Sandwich Quantization Scheme}}}
\vspace*{13mm}

{\textbf{\large{M.M. Sheikh-Jabbari} }}
\vspace*{2mm}

{\it School of Physics, Institute for Research in Fundamental
Sciences (IPM),\\  P.O.Box 19395-5531, Tehran, Iran}\\ \texttt{email:\tcr{jabbari@theory.ipm.ac.ir}}
\vspace*{1cm}
\end{center}
\begin{abstract}
Quantization of field theories with gauge symmetry is an extensively discussed and well-established topic. In this short note, we revisit this old problem. {While we confirm all details of the existing literature, we  highlight a potentially important point which may provide a better understanding of and insights on the quantization of gauged systems}.   The gauge degrees of freedom have vanishing momenta, and hence their equations of motion appear as constraints on the system. We argue that to ensure consistency of quantization  one can impose these constraints as ``sandwich conditions'': The physical Hilbert space of the theory consists of all states for which the constraints sandwiched between any two physical states vanish. We solve the sandwich constraints and show they have solutions not discussed in the gauge field theory literature. We briefly discuss the physical meaning of these solutions and implications of the \textit{sandwich quantization scheme}.

\end{abstract}
\vskip 3mm
\begin{center}
\textit{This is an educational note prepared for the special issue of\\ “Clarifying common misconceptions in high energy physics and cosmology”\\ to be published by Nuclear Physics \textbf{B}.}
\end{center}

\end{titlepage}
\section{Introduction}\label{sec:Intro}

To construct a quantum theory we should apply a ``quantization process'' on a given classical theory. There are some different quantization procedures that for known examples of particle theory and field theory, have been shown to yield the same physical observables. A particularly well-developed formulation is the canonical quantization which is based on working out solution phase space of the classical theory, and promoting the Poisson brackets to commutators and observables to operators. The canonical quantization scheme can be applied to particle theory, yielding standard Heisenberg formulation of quantum mechanics, as well as to quantum field theories (QFT) \cite{Weinberg:1995I}.

In general, one may view particle theory as a  $0+1$ dimensional field theory residing on the particle worldline. Our discussions here hold for a generic $d+1$ dimensional field theory for any $d$, including $d=0$. Thus, in what follows when we say (quantum) field theory, it includes the particle theory as well. In canonical quantization procedure, we should also define Hilbert space of the theory over which the operators act. The Hilbert space is usually constructed based on a ``vacuum state'' and all the other states in the Hilbert space of a (local) QFT are constructed by the action of local operators on the vacuum state \cite{Weinberg:1995I, Peskin:1995ev}. Therefore, by construction, we have a version of  operator-state correspondence (OSC). We note that the OSC introduced here and we use  in this paper need not be a one-to-one correspondence (as is more commonly used and stated in 2d Conformal Field Theories, e.g. see \cite{Polchinski:1998rq}). What we mean by OSC is that all states in the Hilbert space are constructed by the action of at least one operator on a vacuum state.\footnote{We thank Glenn Barnich and Marco Serone for a comment on this point.} (Vacuum state then corresponds to the identity operator.)

Besides the canonical quantization, there is the well-established and commonly used path integral formulation.\footnote{This is despite the fact that a rigorous mathematical definition of the path integral and in particular its measure, is still missing.} There is an extensive literature establishing the canonical quantization and path integral quantization schemes agree on all physical observables. In the path integral formulation of local QFTs we do not have the Hilbert space and the path integral computes all physical observables, that are assumed to be of the form of generic $n$-point functions, vacuum expectation value (VEV) of generic products of local operators \cite{Weinberg:1995I, Peskin:1995ev}. Nevertheless, one can implicitly talk/think about states and Hilbert space of the theory assuming the OSC defined above.

Gauge theories and gauge symmetries have been the cornerstone of physical formulations in the last century. Gauge symmetries, despite the commonly used name, do not point to conserved charges via Noether's first theorem \cite{Weinberg:1995I, Weinberg:1995II}. They are rather ``redundancies of description'' usually introduced to make other symmetries of the system more manifest or as a theoretical guiding principle for fixing interactions among fields.  In particle theory, the gauge symmetry corresponds to reparameterization of the worldline parameter. In the context of field theories, gauge symmetries may correspond to ``internal symmetries'' over the field space, like the case of  Maxwell theory or non-Abelian gauge theories in the standard model of particle physics, or to ``external (spacetime)'' symmetries, like diffeomorphism invariance in gravitational theories \cite{Wald:1984rg}.

Gauge theories are prime examples of constrained systems: Some of the degrees of freedom (d.o.f) in the Lagrangian of gauge theories, the \textit{gauge d.o.f},  have vanishing conjugate momenta.  Thus, equations of motion for the gauge d.o.f. are constraints  and gauge d.o.f are not propagating and dynamical (in contrast to a dynamical equation involving second order time derivatives). As such, systems with gauge symmetry are constrained systems and Dirac's procedure \cite{Dirac:1996} or other suitable methods for quantizing constrained systems should be invoked \cite{Henneaux:1992}.

The standard procedure for dealing with gauge field theories starts with gauge-fixing: If we have $N$ gauge d.o.f, gauge-fixing involves assuming $N$ arbitrary relations among gauge fields and possibly their derivative such that these relations  do not remain invariant under a generic gauge transformation. Thus, fixing these relations amounts to fixing a gauge. Gauge-fixing relations may be viewed as constraints which together with the equations of motion (EoM) for the gauge d.o.f form a well-posed constraint system \cite{Dirac:1996, Henneaux:1992}.\footnote{Note that while the gauge-fixing relations, by construction and definition, are not gauge invariant, equations of motion for the gauge d.o.f,  by the very definition of gauge theory, are \textit{gauge covariant}: the set of EoM is closed under gauge transformations.} One may then construct solution space of the classical gauge-fixed theory and go through the standard canonical quantization procedure. Alternatively, one may choose the path integral method.

In the path integral quantization, one does not use EoM, nonetheless, the gauge-fixing procedure can proceed. The gauge-fixing condition can be inserted into the path integral through a delta-function. This delta-function can be exponentiated using ghosts and the emerging BRST symmetry \cite{Becchi:1975nq, Tyutin:1975qk} guarantees consistency of this procedure, see e.g. \cite{ Peskin:1995ev, Weinberg:1995II,  Henneaux:1992, Itzykson:1980rh}.

In this note, we revisit quantization of gauge field theories, especially the canonical quantization.
We start with a quick but careful review of the standard discussions and observe that EoM of the gauge d.o.f. need not necessarily be imposed as one finds in the standard textbooks; one can impose them as ``sandwich conditions''. That is, physical Hilbert space of the gauge field theory may be defined upon the condition that the EoM of gauge d.o.f,  promoted to operators upon quantization,  vanish when sandwiched between any two physical states.

{Disclaimer: We stress that this note is prepared for a special issue trying to stress simple and basic facts and issues of educational value regarding quantization of gauged systems. By no means, we do not intend to provide a full account of the problem, our goal is to highlight a point, a concept, that has not been noted before and can shed a better light on the gauge systems with potentially interesting and important physical significance, yet to be developed and uncovered. }

\section{Gauge theories, a quick review of basic facts}\label{sec:gauge-theory}

Consider a gauge field theory described by the Lagrangian ${\cal L}(\Phi)$ where $\Phi$ denotes a generic set of fields that includes ``gauge d.o.f.'' $\varphi_i, i=1,2,\cdots, N$ and other fields $\psi_A$. This theory has two important features:
\begin{itemize}
    \item[(1)] Gauge d.o.f $\varphi_i$, by definition, are the fields with identically vanishing conjugate momentum. If $t$ denotes the time direction, that is,\footnote{One should note that while all the gauge d.o.f  have vanishing conjugate momentum, the converse is not necessarily true. For example, consider the Maxwell theory plus a mass term for the vector field, where $A_0$ has a vanishing momentum (cf.  section \ref{sec:Maxwell}), this theory has no gauge symmetry. We thank the anonymous referee for pointing this out.}
    \begin{equation}\label{momentum-varphi}
        \Pi^i:= \frac{\partial {\cal L}}{\partial (\partial_t \varphi_i)}\equiv 0.
    \end{equation}
    \item[(2)] The action is invariant under gauge transformations $\Phi\to \Phi+ \delta_{{\lambda_i}} \Phi$, where $\lambda_i$ denote gauge parameters:
    \begin{equation}\label{gauge-transf-general}
         \delta_{{\lambda_i}}S= \frac{\delta{S}}{\delta \lambda_i}=0 \qquad \text{off-shell},\qquad S:=\int_M\ {\cal L}(\Phi).
    \end{equation}
Note that both $\varphi_i$ and  $\psi_A$ transform under gauge transformations and that number of gauge d.o.f $\varphi_i$ is equal to the number of independent gauge parameters $\lambda_i$.
\end{itemize}
An immediate consequence of the above is that the EoM for gauge d.o.f $\varphi_i$, ${\cal C}^i$, are not dynamical equations; they are constraints:
\begin{equation}\label{gauge-EoM}
          \frac{\delta{S}}{\delta \varphi_i}=0 \quad\Longrightarrow \quad {\cal C}^i:=-\partial_t \Pi^i+\frac{\partial {\cal L}}{\partial  \varphi_i}=\frac{\partial {\cal L}}{\partial  \varphi_i}=0.
    \end{equation}
In the above, we have used the fact that $\Pi^i$ are identically vanishing (at all times). It is important to remember that by the definition of gauge invariance \eqref{gauge-transf-general} (which should hold off-shell),  field equations are gauge covariant (their set is gauge invariant). So, the constraints ${\cal C}^i$ are gauge covariant and their set closes onto itself under gauge transformations.

\subsection{Gauge-fixing}

Given the freedom in the choice of $N$ gauge parameters $\lambda_i$, we can fix $N$ generic combinations of the fields $\Phi$ and set it to zero,
\begin{equation}\label{gauge-fixing-G}
    {\cal G}_i (\varphi_i,\psi_A)=0\,.
\end{equation}
${\cal G}_i$ specify the desired gauge-fixing of our choice and are functionals not invariant under gauge transformations. In other words, to have a consistent gauge-fixing $\det\left(\frac{\delta {\cal G}_i}{\delta \lambda_j}\right)\neq 0$ where ${\frac{\delta {\cal G}_i}{\delta \lambda_j}=\frac{\delta {\cal G}_i}{\delta\Phi}{\delta}_{\lambda_j}\Phi}$.
As such, \eqref{gauge-fixing-G} fixes the freedom in choices of $\lambda_i$: under a gauge transformation ${\cal G}_i=0$ will not hold and requiring \eqref{gauge-fixing-G} amounts to specific choice for $\lambda_i$.

Consistency of the gauge-fixing requires that \eqref{gauge-fixing-G}  is compatible with the evolution of the system \cite{Henneaux:1992}, i.e. it should hold at all times, in particular, ${\cal G}_i=0$ and ${\cal C}^i=0$ should hold simultaneously. This consistency requirement may be analyzed and carried out  in the Hamiltonian formulation, yielding  the chain of secondary constraints \cite{Henneaux:1992} or, in the Lagrangian (action) formulation \cite{Weinberg:1995I,Peskin:1995ev, Itzykson:1980rh}, yielding physically identical results. A particularly simple and handy gauge-fixing, that we adopt in the rest of this work, is to fix the functional form of gauge d.of. as
\begin{equation}\label{our-gauge-fixing}
    \varphi_i-\varphi_i^0=0,
\end{equation}
where $\varphi^0_i$ is a given function with no dependence on other fields; temporal or axial gauge-fixing is an example of the gauge-fixing in \eqref{our-gauge-fixing}.

The classically physical (gauge inequivalent) field configurations are hence configurations of $\psi_A$ subject to their EoM with $\varphi_i-\varphi_i^0=0$ and ${\cal C}^i=0$ (at all times). As we will discuss in more detail in section \ref{sec:Maxwell}, for the case of Maxwell theory this physical solution space is described by transverse polarization of electromagnetic waves (with $d-2$ polarizations for the $d$ dimensional theory) plus the boundary soft modes specified by functions on codimension 2 celestial sphere. See \cite{Strominger:2017zoo} for more details.

\subsection{Right-action quantization scheme}

Having the classical physical solution space, one may proceed with the canonical quantization, i.e. one may promote the physical field configurations and their momenta to operators that satisfy the canonical commutation relations. This guarantees the expectation that one-particle Hilbert space of the quantum theory and the classical field configurations are in one-to-one relation. This is the standard procedure introduced by Julian Schwinger in 1948 \cite{Schwinger:1948yj, Schwinger:1948yk} and developed further in 1950 by Gupta \cite{Gupta:1949rh} and Bleuler \cite{Bleuler:1950cy}. See \cite{Itzykson:1980rh} for a review of the Gupta-Bleuler method which is a well-known but old and somewhat outdated quantization method. The Gupta-Bleuler method was surpassed by the BRST method \cite{ Henneaux:1992, Becchi:1975nq, Tyutin:1975qk}.

In what follows, the operator associated with a generic function(al) of fields ${\cal O}[\Phi]$ will be denoted by $\bcO$, the vector space of all field configurations by $\cHt$ and the Hilbert space of all physical field configurations by $\cHp$. One can construct $\cHt$ as in standard QFT textbooks:
{\begin{enumerate}
    \item Construct solutions to the classical EoM;
    \item Promote the ``Fourier coefficients'' of the solutions to the canonical creation/annihilation operators or to the positive and negative frequency modes.
    \item Define the vacuum state $|0\rangle$ as the state annihilated by all annihilation operators.
    \item Construct one-particle sector of $\cHt$  by the action of the creation operators on the vacuum state $|0\rangle$. The whole $\cHt$ is the Fock space built upon this one-particle vector space.
    \item Note that for gauge theories, generically, $\cHt$ is not a Hilbert space, because it contains zero or negative norm states. After removing these states (using ghosts or other methods), we obtain the physical Hilbert space $\cHp$ which only contains positive and finite norm states; physical states may be normalized to have norm 1.
\end{enumerate}
The main point of the gauge-fixing procedure is item 5. above and removing the unphysical states/d.o.f.\  In the now-standard textbook treatment, removal of these states is performed through ghosts and is guaranteed by the BRST symmetry \cite{Peskin:1995ev, Weinberg:1995II,  Henneaux:1992, Itzykson:1980rh}. Here, we will not take the route of ghosts and the BRST; for illustrative purposes of our main idea, we follow axial gauge-fixing in which it is known that the ghosts decouple and we revisit Gupta-Bleuler procedure.}

In the adopted notation, functionals associated with \eqref{our-gauge-fixing} and the EoM of $\varphi_i$  will be respectively denoted by $\boldsymbol{\varphi}_i-\boldsymbol{\varphi}^0_i$ and  $\boldsymbol{{\cal C}}^i$ and, the positive frequency sector of these, {the part consisting of creation operators,} will be denoted by $\left(\boldsymbol{\varphi}_i-\boldsymbol{\varphi}^0_i\right)^+$ and  $\left(\boldsymbol{{\cal C}}^i\right)^+$.
    When we discuss vanishing of an operator we should specify over which vector/Hilbert space it vanishes. In our case, $\boldsymbol{\varphi}_i-\boldsymbol{\varphi}^0_i$ and $\boldsymbol{{\cal C}}^i$ (or the positive frequency parts of them) do not vanish over $\cHt$ but are zero over $\cHp$. To be precise, the Schwinger-Gupta-Bleuler quantization scheme \cite{Schwinger:1948yj, Schwinger:1948yk, Gupta:1949rh, Bleuler:1950cy}, or the ``right-action quantization scheme'' requires\footnote{Since the positive and negative frequency parts of an operator by construction do not commute with each other, it is clear that one cannot require $\left(\boldsymbol{\varphi}_i-\boldsymbol{\varphi}^0_i\right)^+|\psi\rangle=0$ and $\left(\boldsymbol{\varphi}_i-\boldsymbol{\varphi}^0_i\right)^-|\psi\rangle=0$ simultaneously.}
\begin{equation}\label{RA-constraints}
    \left(\boldsymbol{\varphi}_i-\boldsymbol{\varphi}^0_i\right)^+|\psi\rangle=0, \qquad \left(\boldsymbol{{\cal C}}^i\right)^+|\psi\rangle=0, \quad \forall\ |\psi\rangle \in \cHp.
\end{equation}
The above is basically what we can find in old QFT textbooks, e.g. \cite{Itzykson:1980rh} and is shown to yield a consistent quantization scheme for gauge theories. We call this the ``right-action quantization scheme'', as the right-action of the positive-frequency part of the constraint operators on the physical Hilbert space is required to vanish.

The question we would like to explore more carefully is whether the right-action quantization scheme is a necessary requirement or passage from classical physical (gauge inequivalent) field configurations to physical quantum Hilbert space can be made with a weaker requirement. We argue in the next section that indeed a weaker condition can fulfill the requirement.

\section{Sandwich quantization scheme}\label{sec:SQS}

We start with recalling that operators on a vector space $\cHt$ may customarily be viewed as matrices over the vector space, i.e. a given operator $\boldsymbol{{\cal O}}$ over $\cHt$ may be replaced with $\langle\psi|\bcO|\phi\rangle$ for any two states $|\phi\rangle, |\psi\rangle\in \cHt$. So, a natural choice/proposal is to replace vanishing of an operator by vanishing of its matrix elements. Explicitly, one may explore replacing \eqref{RA-constraints} with the ``sandwich constraints'':
\begin{equation}\label{Sandwich-constraints-1}
    \langle\phi|(\boldsymbol{\varphi}_i-\boldsymbol{\varphi}^0_i)|\psi\rangle=0, \qquad \langle\phi|\boldsymbol{{\cal C}}^i|\psi\rangle=0, \quad \forall\ |\psi\rangle, |\phi\rangle \in \cHp.
\end{equation}
In other words, one may define physical Hilbert space  $\cHp$ upon \eqref{Sandwich-constraints-1}. All solutions to \eqref{RA-constraints} are also solutions to \eqref{Sandwich-constraints-1}, but the reverse is not necessarily true.

One can adopt a quantization scheme using a mixture of sandwich and right-action conditions: impose gauge-fixing condition at classical level and impose $\bcC^i$  as a sandwich constraint. We adopt this quantization scheme, explicitly:
\begin{equation}\label{Our-quatization-scheme}
 \tcbset{fonttitle=\scriptsize}
            \tcboxmath[colback=white,colframe=gray]{\hspace*{-4mm} {\varphi}_i-{\varphi}^0_i=0, \ \ \langle \psi'|  \bcC^i |\psi\rangle=0, \ \ \forall |\psi\rangle, |\psi'\rangle \in \cHp \hspace*{-4mm}}
\end{equation}
{To proceed, one should first construct the classical solution space obtained in the (axial-type) ${\varphi}_i-{\varphi}^0_i=0$ gauge. Then, one should quantize the solution space and construct the $\cHt$ by promoting these solutions to operators acting on vacuum state (which is defined as the state eliminated by annihilation operators).} The physical Hilbert space $\cHp$ is then a subset of $\cHt$ that satisfies \eqref{Our-quatization-scheme}. We argue in the next subsection that the above sandwich quantization scheme \eqref{Our-quatization-scheme} may be obtained via path integral quantization.

We close this part by the remark that it was already noted in the old literature \cite{Gupta:1949rh, Bleuler:1950cy} (see also \cite{Kowalski-Glikman:1990pnu, Kowalski-Glikman:1992bnv})  that \eqref{Sandwich-constraints-1} are  what one physically needs for quantization. However, in the Gupta-Bleuler quantization only the gauge-fixing condition was required to hold as the sandwich condition (the left equation in \eqref{Sandwich-constraints-1}) and in practice, it was imposed and solved by the right-action requirement; the EoM for the gauge d.o.f $\bcC^i=0$ was not considered.

\subsection{Path integral derivation of sandwich quantization scheme \texorpdfstring{\eqref{Our-quatization-scheme}}{ref}}\label{sec:derivation}

Let $\bcO_i(x)$ denote  \textit{gauge invariant} local operators in the gauge field theory we study; that is, by definition,
\begin{equation}\label{gauge-invariance}
    \delta_{\lambda_j} {\cal O}_i=\frac{\delta {\cal O}_i[\Phi]}{\delta \Phi} \delta_{\lambda_j} \Phi= 0.
\end{equation}
Physical observables are then generic $n$-point functions of these operators, VEV of time-ordered product of these operators:\footnote{We are assuming that vacuum state $|0\rangle$ is a state in $\cHt$.}
\begin{equation}
    G_n(x_1,x_2,\cdots, x_n)= \langle \bcO_1(x_1) \bcO_2(x_2)\cdots \bcO_n(x_n)\rangle.
\end{equation}
The above $n$-point function, which is gauge invariant by definition,  may be computed using the path integral,
\begin{equation}
    G_n(x_1,x_2,\cdots, x_n)= \int {\cal D}\Phi \ {\cal O}_1(x_1) {\cal O}_2(x_2)\cdots {\cal O}_n(x_n) \ e^{\frac{i}{\hbar} S[\Phi]}
\end{equation}
To perform the above path integral, one should either define the measure ${\cal D}\Phi$ by modding it out by the volume of the gauge orbits or equivalently, insert the gauge-fixing condition \cite{Peskin:1995ev, Weinberg:1995II, Itzykson:1980rh}, in our case $\varphi_i-\varphi_i^0=0$. Let us denote the modded out measure by $\overline{{\cal D}\Phi}$ and let ${\bar S}$ be the action computed at $\varphi_i-\varphi_i^0=0$, i.e. $\bar S= S(\varphi_i^0, \psi_A)$.\footnote{Note that in a more general gauge ${\cal G}_i=0$ \eqref{gauge-fixing-G}, one should insert the Jacobian of transformation, $\det(\delta {\cal G}_i/\delta \lambda_j)$. Exponentiating this Jacobian using ghost fields $c_i, \bar{c}^i$, the total (gauge-fixed plus ghost) action is $S_{\text{tot}}=S+ X^i{\cal G}_i+ \bar{c}^i \left(\frac{\delta {\cal G}_i}{\delta \lambda_j}\right) c_j$ where $X^i$ are Langrange multipliers. The total action then exhibits the BRST symmetry \cite{Barnich:2000zw}.} In the axial-type gauge we adopted here, ghosts decouple and we need not consider them.

To study the constraints at quantum level in the path integral formulation, we can/should consider a generic $n$-point function with the insertion of ${\cal C}^i$. This is a very common analysis in the study of anomalies and gauge-fixings in gauge field theories \cite{Peskin:1995ev, Weinberg:1995II} and/or string theory, see in particular the first few chapters in \cite{Polchinski:1998rq}.
\begin{equation}\begin{split}
    \hspace*{-5mm}\langle \bcO_1(x_1) \bcO_2(x_2)\ {\bcC}^i(x) \cdots \bcO_n(x_n)\rangle &:=\int \overline{{\cal D}\Phi} \ {\cal O}_1(x_1) {\cal O}_2(x_2)\cdots {\cal O}_n(x_n)\ {\cal C}^i(x)\ e^{\frac{i}{\hbar} \bar{S}[\Phi]}\\
    &=-i{\hbar}\int \overline{{\cal D}\Phi} \ {\cal O}_1(x_1) {\cal O}_2(x_2)\cdots {\cal O}_n(x_n)\ \left(\frac{\delta}{\delta\varphi_i(x)} e^{\frac{i}{\hbar} S[\Phi]}\right)_{\varphi_i=\varphi_i^0}\\
    &=i{\hbar}\int \overline{{\cal D}\Phi} \ \frac{\delta}{\delta\varphi_i(x)}\left({\cal O}_1(x_1) {\cal O}_2(x_2)\cdots {\cal O}_n(x_n)\right)\big{|}_{\varphi_i=\varphi_i^0}\  e^{\frac{i}{\hbar} \bar{S}[\Phi]}
\end{split}
\end{equation}
where in the second line we used the fact that ${\cal C}^i$ are EoM for the gauge d.o.f $\varphi_i$ \eqref{gauge-EoM} and in the third line we used integration by-part.

To proceed further,  recall that one may replace $\varphi_i$ with the gauge parameters, i.e. $\varphi_i=\varphi_i[\lambda_j]$ and as such,
\begin{equation}\label{n-point-Ca-2}
     \langle \bcO_1(x_1) \cdots\ {\bcC}^i(x) \cdots \bcO_n(x_n)\rangle =i{\hbar}\int \overline{{\cal D}\Phi} \ \frac{\delta\lambda_j(y)}{\delta\varphi_i(x)}\ \delta_{\lambda_j(y)}\bigg({\cal O}_1(x_1)\cdots {\cal O}_n(x_n)\bigg)\  e^{\frac{i}{\hbar} S[\Phi]}
\end{equation}
Noting that ${\bcO}_i$ are gauge-invariant local operators \eqref{gauge-invariance}, we learn that
\begin{equation}\label{n-point-Ca}
    \tcboxmath[colback=white,colframe=gray]{\hspace*{-4mm}  \langle \bcO_1(x_1) \cdots\ {\bcC}^i(x) \cdots \bcO_n(x_n)\rangle =0 \qquad \forall\ \text{local gauge-invariant operators}\ {\bcO}_i.\hspace*{-4mm}}
\end{equation}

One may recall the OSC defined in the introduction, that to any local gauge-invariant  operator ${\bcO}_i$ one may associate a state, conveniently denoted by $|{\cal O}_i\rangle$. The vacuum state $|0\rangle$ corresponds to the identity operator and a generic (multi-particle) state may be associated with products of such operators. In this terminology, \eqref{n-point-Ca} is equivalent to sandwich constraints \eqref{Our-quatization-scheme}. In other words, we have presented a derivation of the sandwich quantization scheme \eqref{Our-quatization-scheme} from the path integral formulation. This derivation also implies that if
$\bcO_1$ and $\bcO_2$ are two physical operators ($|{\cal O}_1\rangle, |{\cal O}_2\rangle \in \cHp$), then $\bcO_1 \bcO_2 |0\rangle$ should also be in $\cHp$; see appendix B of \cite{Dutta:2024gkc} for related arguments.

Before moving on and discussing the proposal for solving the sandwich constraints, we remark that the right-hand-side of \eqref{n-point-Ca-2} is proportional to $\hbar$. At the quantum level, one may have defined the constraints as $\frac{1}{\hbar}\bcC^i$, making more manifest that the sandwich quantization scheme is an option appearing at the quantum level, while at the classical level we deal with ${\cal C}^i=0$.

\subsection{Solving sandwich conditions}\label{sec:Solving-sandwich-condition}

To solve \eqref{Our-quatization-scheme}, we first employ the ideas developed in \cite{Dutta:2024gkc, Bagchi:2024tyq} and then use the construction for the case of a field theory.

\paragraph{General idea of the solution.} Consider a Hermitian operator $\bcC$  that acts on the total vector space $\cHt$ with the requirement that all states in $\cHp\subset \cHt$ should satisfy $\langle \psi'|  \bcC |\psi\rangle=0$. Let $|C\rangle$ denote eigenstates of $\bcC$,
\begin{equation}\label{C-eigenstates}
    \bcC |C\rangle ={C} |C\rangle.
\end{equation}
Zero-eigenstates with ${C}=0$ are physical states; however,  the sandwich condition $\langle \psi'|  \bcC |\psi\rangle=0$ allows for other solutions with ${C}\neq 0$.

Assume that $\bcC$ has a $\mathbb{Z}_2$ symmetric spectrum such that for every state with eigenvalue $|{C}|$ there is another state with eigenvalue $-|{C}|$. That is, assume there exists a (anti)Hermitian operator  \textbf{\sho} such that
\begin{equation}
\textbf{\sho}\bcC\textbf{\sho}=-\bcC, \qquad \textbf{\sho}^2=1, \qquad  \textbf{\sho}|C\rangle= |-C\rangle\,.
\end{equation}
One can construct states of the form
\begin{equation}
    |\mathtt{C}\rangle_{\pm}:=\frac{1}{\sqrt{2}}(|\mathtt{C}\rangle \pm  |-\mathtt{C}\rangle)=\frac{1}{\sqrt{2}}(1\pm \textbf{\sho})|\mathtt{C}\rangle  ,\qquad \mathtt{C}:=|{ C}|
\end{equation}
For these states we find:
\begin{equation}
\bcC |\mathtt{C}\rangle_{\pm}= \mathtt{C}\  |\mathtt{C}\rangle_{\mp}, \qquad {}_{\mp}\langle \mathtt{C}'|\mathtt{C}\rangle_{\pm}=0,
\end{equation}
and ${}_{\pm}\langle \mathtt{C}'|\mathtt{C}\rangle_{\pm}=\delta_{\mathtt{C}, \mathtt{C}'} \ (\text{for} \ \mathtt{C}\neq 0)$. Thus, evidently ${}_{\pm}\langle \mathtt{C}'|\bcC |\mathtt{C}\rangle_{\pm}=0$ for any $\mathtt{C}, \mathtt{C}'$. Consequently, the collection of all states $|\mathtt{C}\rangle_{+}$ (or $|\mathtt{C}\rangle_{-}$) form a physical Hilbert space. To include $C=0$ solutions in the same convention, we choose $\cHp$  to be spanned by  $|\mathtt{C}\rangle_{+}$ states. Note that the $\mathtt{C}=0$ sector of $\cHp$ respects the $\mathbb{Z}_2$ and generic $\mathtt{C}\neq 0$ does not exhibit this symmetry.

With the above observation and conventions, we are now ready to classify solutions to \eqref{Our-quatization-scheme}. We have two options,
\begin{equation}\label{Two-classes}
\begin{split}
&\textbf{(A):}\qquad   \bcC |\psi\rangle \in \cHp, \ \forall |\psi\rangle \in \cHp, \\
&\textbf{(B):}\qquad  \bcC |\psi\rangle \in \cHc, \ \forall |\psi\rangle \in \cHp, \quad \cHc:=\cHt -\cHp.
\end{split}
\end{equation}
That is, for the case (A), $\bcC$ is an operator defined over the physical Hilbert space $\cHp$ and hence \eqref{Our-quatization-scheme} implies that physical states must be zero eigenstates of $\bcC$. For (B), however,  $\bcC$ is defined over $\cHt$ and not $\cHp$. In our notation, $\cHt$ is divided into two parts, $\cHp$ and its complement $\cHc$: $\cHt=\cHp\cup \cHc$. Without loss of generality one can choose
\begin{equation}\label{HpHc-orthogonality}
   \langle \psi_{\text{c}} |\psi_{\text{p}}\rangle =0, \quad \forall |\psi_{\text{p}}\rangle \in \cHp, \ |\psi_{\text{c}}\rangle \in \cHc .
\end{equation}
Note that within our construction,  \eqref{HpHc-orthogonality} is an immediate outcome and we need not an independent proof:
If $\cHp$ is spanned by $|\mathtt{C}\rangle_{+}$, then $\cHc$ is spanned by $|\mathtt{C}\rangle_{-}$, which are by construction orthogonal sets.\footnote{We thank Marco Serone for a question in this regard.}

\paragraph{Application to the case of a gauge field theory.} {The above was a general construction which works well (without extra mathematical subtleties) for finite or non-continuous infinite-dimensional Hilbert spaces. For the case of quantum field theories, where states in $\cHt$ have a continuous label, one should be more careful with some mathematical subtleties. Here, we do not intend to delve into and address all those mathematical subtleties. They need a more in-depth study and will distract the reader from the main conceptual points we intend to raise and discuss in this short and educational note. They will be reported upon in future publications. The first issue is with \eqref{C-eigenstates}. One should note that for a generic operator-function $\bcC$, \eqref{C-eigenstates} is only meaningful if $\bcC$ at different points in space commute with each other.\footnote{We thank Sergio Cecotti for discussion on this issue.} In field theories where we usually make a perturbative analysis, one can work with Fourier transform of the constraint $\bcC$, and this issue requires commuting constraints at different values of momenta. Moreover, as in the usual Schwinger-Gupta-Bleuler method, we only need to consider \eqref{C-eigenstates} for positive frequency modes or in cases like that of string theory \cite{Bagchi:2024tyq} where we deal with constraints quadratic in creation-annihilation operators, we consider the normal ordered constraints. More on discussions on these subtleties will be given below. For now and with these precautions and just to illustrate the basic idea, let us employ the above $\mathbb{Z}_2$ based construction, to solve \eqref{Our-quatization-scheme}}:
\begin{equation}\label{Two-classes-soln}
\begin{split}
&\textbf{Class 1:}\qquad   (\bcC)^+ |\psi\rangle=0, \qquad |\psi\rangle \in \{|\mathtt{C}=0\rangle_{+} \}\\
&\textbf{Class 2:}\quad   \bcC |\psi\rangle\in\cHc, \quad (\bcC)^+ |\psi\rangle \neq 0, \qquad |\psi\rangle \in \{|\mathtt{C}\rangle_{+},\ \mathtt{C}\neq 0 \}.
\end{split}
\end{equation}
That is, Class 1 Hilbert space is the part of $\cHt$ that is invariant under \textbf{\sho}, whereas Class 2 states under \textbf{\sho} are mapped onto states in $\cHc$.
In other words, $\bcC$ maps a Class 2 physical state onto a state in the complement Hilbert space $\cHc$ while $\bcC^2$ takes a physical Class 2 state to a physical state. As pointed out in \cite{Dutta:2024gkc}, Class 1 and Class 2 are two super-selection sectors in the physical Hilbert space defined through the quantization scheme in \eqref{Our-quatization-scheme}. There is no overlap between the Class 1 and Class 2 Hilbert spaces and the Class 1 Hilbert space is a part of $\cHc$ as seen by the Class 2 physical Hilbert space.

We close this section with some important remarks:
\begin{enumerate}
    \item The vacuum state of the total Hilbert space $\cHt$ is, by definition, a state in the Class 1 Hilbert space, and hence \textit{cannot} be a state in the Class 2 Hilbert space.
    \item {The difference between Class 1 and Class 2 Hilbert spaces is in the vacuum state upon which they are built. Explicitly, as is textbook material, all states in the Class 1 physical Hilbert space can be constructed by the action of all gauge-invariant local operators on the Class 1 vacuum state. In the OSC terminology, the Class 1 vacuum state corresponds to the identity operator in this physical Hilbert space.}
    \item {In a similar way, the Class 2 physical Hilbert space is constructed by the action of the same gauge-invariant local operators on the Class 2 vacuum state.}
    \item {However, the crucial fact to note is that Class 2 Hilbert space consists of infinitely many (a continuum for a generic gauge field theory) of super-selection sectors. That is, we are not dealing with a single Class 2 vacuum state or Class 2 physical Hilbert space; we have a continuum of Class 2 vacuum states or physical Hilbert spaces.}
    \item {To understand the above comment, recall the discussion above \eqref{Two-classes-soln} and that the constraint is an operator on the spacetime. One can require the Class 2 vacuum state corresponds to a non-vanishing Fourier mode of the constraint for a given momentum. We will make this a bit more explicit in the example we discuss in the next section.}
    \item {In a different wording, the difference between different physical Hilbert spaces, Class 1 or Class 2 cases, is that they are built upon different vacuum states. Each super-selection sector has its own ``identity operator,'' nonetheless, these Hilbert spaces are in one-to-one correspondence and the consistency of construction requires that physics built upon either of these Hilbert spaces should be the same. So, working with the textbook choice of Class 1 vacuum and Hilbert space is enough to describe physics.}
\end{enumerate}

\section{Example: Maxwell theory}\label{sec:Maxwell}

As a simple but illustrative example, we work through constructing the Class 2 states for the $d$ dimensional Maxwell theory, where $\Phi$ are the gauge field components $A_\mu$ with the Lagrangian
\begin{equation}\label{Maxwell-action}
    {\cal L}=-\frac{1}{4} \ F_{\mu\nu}F^{\mu\nu},\qquad F_{\mu\nu}:=\partial_\mu A_\nu-\partial_\nu A_\mu.
\end{equation}
Equations of motion are of the form
\begin{equation}
    \partial^\mu\partial_\mu A_\nu- \partial_\nu (\partial^\mu A_\mu) =0,
\end{equation}
and most general solutions are,
\begin{equation}\label{generic-soln-Maxwell}
      A_\mu=\int \text{d}^dk\ \left(\epsilon_\mu(k) \delta(k^2) + \widehat{A(k)} k_\mu\right) e^{-i(\omega t+\vec{k}\cdot \vec{x})}+c.c.,\qquad k^\mu \epsilon_\mu(k)=0,\ k^2=-\omega^2+|\vec{k}|^2
\end{equation}
where $\epsilon_\mu(k)$ is a vector subject to $k^\mu \epsilon_\mu(k)=0$ and $\widehat{A(k)}$ is an arbitrary function of $k_\mu$.

The conjugate momenta to $A_\mu$ are
\begin{equation}
    \Pi^\mu= \frac{\partial {\cal L}}{\partial (\partial_t A_\mu)}= F^{t\mu}.
\end{equation}
As we see $\Pi^t$ identically vanishes and hence $A_t$ is the gauge d.o.f. So, in our notation, $\varphi_i$ is $A_t$ and $\psi_A$ are the spatial components of the gauge field $A_a$ and under generic gauge transformations, $A_\mu\to A_\mu+\partial_\mu\lambda$. Note that both $A_t$ and $A_a$ transform under a gauge transformation. There are various gauges used in the literature \cite{Peskin:1995ev, Henneaux:1992}, e.g.  Lorenz gauge $\partial^\mu A_\mu=0$ or temporal gauge $A_t=\varphi_0(\vec{x})$ for a given function $\varphi_0(\vec{x})$. Here we adopt the latter, yielding
\begin{equation}
    \widehat{A(k)}= A(k) -\frac{1}{\omega}\epsilon_t(k) \delta(k^2), \qquad 2\int\ \text{d}^dk\ \omega\ \text{Re}\left(A(k) e^{-i(\omega t+\vec{k}\cdot \vec{x})}\right)=\varphi_0(\vec{x}).
\end{equation}
A generic solution for $A(k)$ is,
\begin{equation}
    A(k) = \left(\tilde{A}(\vec{k})+\frac{1}{\omega}\varphi_0(\vec{k})\right) \delta(\omega),
\end{equation}
where $\varphi_0(\vec{k})$ and $\tilde{A}(\vec{k})$ are respectively  Fourier modes of $\varphi_0(\vec{x})$ and an arbitrary (unspecified) function $\tilde{A}(\vec{x})$.
Thus, the solution \eqref{generic-soln-Maxwell} in this gauge takes the form
\begin{equation}\label{temporal-gauge-soln-Maxwell}
    A_a=\int \text{d}^dk\ \bigg(\epsilon^\perp_a(k) \delta(k^2) + A(\vec{k}) k_a \bigg) e^{-i(\omega t +\vec{k}\cdot \vec{x})}+c.c.,
\end{equation}
where
\begin{equation}
    \qquad \epsilon_t(k)=\frac{1}{\omega} k^a\epsilon_a(k),\qquad \epsilon^\perp_a(k):=\epsilon_a(k)-k^b\epsilon_b  \frac{k_a}{|\vec{k}|^2}
\end{equation}
With the above, we have
\begin{equation}\label{Electric-field-soln}
\begin{split}
    E_a =\partial_t A_a-\partial_a\varphi_0(\vec{x}) &=2\int \text{d}^dk\ \omega\ \delta(k^2)\ \text{Im}\big(\epsilon^\perp_a(k)  e^{-i(\omega t+\vec{k}\cdot \vec{x})}\big)-\partial_a\varphi_0(\vec{x}),\\ 
  \vec{\nabla}\cdot\vec{E}&= 
  -\vec{\nabla}^2\varphi_0(\vec{x}).
\end{split}
\end{equation}
Note that the consistency of Maxwell's equations implies
\begin{equation}
\partial_t(\vec{\nabla}\cdot\vec{E})=0 \qquad \text{or} \qquad \vec{\nabla}\cdot\vec{E}\ \text{should be constant in time}.
\end{equation}
The constraint (EoM for $A_t$) is the Gauss law: $\vec{\nabla}\cdot \vec{E}=0$, $E^a=F^{ta}$ is the electric field strength, the momentum conjugate to $A_a$. In this case and since we are dealing with an Abelian gauge theory,  the constraint is gauge-invariant (instead of being covariant).

Solutions to the classical equations of motion after fixing $A_t=\varphi_0(\vec{x})$ gauge and imposing the constraint $\vec{\nabla}\cdot \vec{E}=0$ is
\begin{equation}\label{classical-soln-space-Maxwell}
    A_a=A_a^\perp(t;\vec{x})+\partial_a \tilde A(\vec{x}), \qquad A_t=\varphi_0(\vec{x}),\quad  \vec{\nabla}^2\varphi_0(\vec{x})=0,
\end{equation}
where $A_a^\perp(t;\vec{x})$ denotes transverse ingoing or outgoing electromagnetic waves, $\varphi_0(\vec{x})$ is the standard Coulomb potential and $\tilde A(\vec{x})$ denotes the longitudinal \textit{soft} modes ($\omega=0$). The latter is time independent and non-dynamical but is crucial to remove infrared (IR) divergences, see \cite{Peskin:1995ev, Weinberg:1995II} and also more recent literature \cite{Strominger:2017zoo}. To quantize the system, as in the standard textbooks \cite{Peskin:1995ev, Itzykson:1980rh}, we promote the Fourier coefficients $\epsilon^\perp_a(k), \varphi_0(\vec{k})$ to operators and impose canonical commutation relations.

To work through the sandwich quantization procedure we outlined above and stated in \eqref{Our-quatization-scheme}, instead of \eqref{classical-soln-space-Maxwell}, we start with \eqref{temporal-gauge-soln-Maxwell}, that is a solution to Maxwell's field equations after fixing the temporal gauge, but before imposing $\vec{\nabla}\cdot \vec{E}=0$ constraint. The solution \eqref{temporal-gauge-soln-Maxwell} is specified by four functions $\epsilon^\perp_a(k), \varphi_0(\vec{k}), \tilde{A}(\vec{k})$. Since the soft part proportional to $\tilde{A}(\vec{k})$ will not be relevant to our discussions, we will drop that part and focus only on the part proportional to $\epsilon^\perp_a(k), \varphi_0(\vec{k})$. We will return to this point in the end of this section.

The total vector space is  obtained by the action of $(\boldsymbol{\epsilon}^\perp_a(k))^\dagger, \boldsymbol{\varphi_0}^\dagger(\vec{k})$ on the vacuum state. The constraint is now
$\vec{\nabla}\cdot \vec{\mathbf{E}}=0$. Its spectrum exhibits the  $\mathbb{Z}_2$ symmetry we required in our construction: the operator \textbf{\sho} is the charge conjugation.  So, we can simply work through the general procedure outlined in the previous section. One readily sees that Class 1 states  are the usual transverse photon states we find in QED.

Class 2 states are gauge field configurations that satisfy the constraint as sandwich conditions:
\begin{equation}\label{Maxwell-sandwich-2}
    \langle\psi|\boldsymbol{\mathcal{Q_B}}|\phi\rangle =0,\qquad (\boldsymbol{\mathcal{Q_B}})^+ |\phi\rangle\neq 0,\qquad \forall  |\psi\rangle,  |\phi\rangle \in \cHp.
\end{equation}
where
\begin{equation}
    \boldsymbol{\mathcal{Q_B}}:= \vec{\nabla}\cdot \vec{\mathbf{E}}
\end{equation}
is the \textit{observer's (background) electric charge density operator}.
For explicit construction of Class 2 states, we use eigenstates of the Hermitian operator $\boldsymbol{\mathcal{Q_B}}$.
Consider eigenstates of $\boldsymbol{\mathcal{Q_B}}$:
\begin{equation}
    \boldsymbol{\mathcal{Q_B}} |\mathcal{Q_B},\pm \rangle=\pm \mathcal{Q_B}|\mathcal{Q_B},\pm\rangle, \qquad \textbf{\sho}|\mathcal{Q_B},\pm\rangle=\mp|\mathcal{Q_B},\pm\rangle 
\end{equation}
{We pause to discuss one the mathematical subtleties mentioned in the previous section in the paragraph above \eqref{Two-classes-soln}, for the example at hand. In this case $\boldsymbol{\mathcal{Q_B}}=\boldsymbol{\mathcal{Q_B}}(x)$ (and not $t$). Since $\boldsymbol{\mathcal{Q_B}}$ is made out of electric field strength $E$ which are momentum conjugate to $A_a$, one can readily observes that $[\boldsymbol{\mathcal{Q_B}}(x), \boldsymbol{\mathcal{Q_B}} (x')]=0$. So one can diagonalize the operator-valued function at different points.} We note that  under charge conjugation \textbf{\sho}, $\boldsymbol{\mathcal{Q_B}}\to \textbf{\sho} \boldsymbol{\mathcal{Q_B}} \textbf{\sho}= -\boldsymbol{\mathcal{Q_B}}$. Therefore, states of the form
\begin{equation}\label{Q-Class-2}
    |\mathcal{Q_B}\rangle_+ :=\frac{1}{\sqrt{2}}(|\mathcal{Q_B},+ \rangle+|\mathcal{Q_B},- \rangle), \qquad \textbf{\sho}|\mathcal{Q_B}\rangle_+=\frac{1}{\sqrt{2}}(|\mathcal{Q_B},+ \rangle - |\mathcal{Q_B},- \rangle)
\end{equation}
solve the constraint equation. Here $\mathcal{Q_B}=\mathcal{Q_B}(\vec{x})$ is generic function up to the constraint that $\mathcal{Q_B}$ and $-\mathcal{Q_B}$ are not viewed as independent. $\mathcal{Q_B}=0$ sector is special, as in this case \textbf{\sho}$|0\rangle_+= 0$; it produces Class 1 states. $\mathcal{Q_B}\neq 0$ produce Class 2 states.

We close this section with some comments and discussions.
\begin{itemize}
\item[(1)] Note that we are considering Maxwell's theory in the absence of electromagnetic genuine physical current and $\boldsymbol{\mathcal{Q_B}}$ should not be viewed as a genuine (physical) electric charge density operator. We are requiring vanishing of $\boldsymbol{\mathcal{Q_B}}$ through sandwich conditions; $\langle\psi'|\boldsymbol{\mathcal{Q_B}}|\psi\rangle=0$ for any two physical states $|\psi\rangle, |\psi'\rangle$.

\item[(2)] We crucially note again that $\boldsymbol{\mathcal{Q_B}}$ on any Class 2 physical state is not a physical state, it is in the complement part $\cHc$.

\item[(3)]  Any state of the form   \eqref{Q-Class-2} for any $\mathcal{Q_B}(\vec{x})$ should be viewed as a \textit{vacuum state} of Class 1 or Class 2 sector. This vacuum state may be associated with an \textit{observer} with a non-trivial background charge density.  That is, depending on the choice of the observer (specified through $\mathcal{Q_B}(\vec{x})$) the physical Hilbert space consists of (infinitely) many super-selection sectors, each specified by a given $\mathcal{Q_B}(\vec{x})$.

\item[(4)] All the other excitations above a given vacuum state may be constructed by the action of transverse polarization creation operators, $(\boldsymbol{\epsilon}^\perp_a)^\dagger(k)$, on either of  vacuum states \eqref{Q-Class-2} with a given $\mathcal{Q_B}(\vec{x})$.  Different super-selection sectors in the physical Hilbert space can correspond to different physical observers which carry multipole charge densities specified through $\mathcal{Q_B}(\vec{x})$.

\item[(5)] For generic $\mathcal{Q_B}(\vec{x})$ in \eqref{Q-Class-2}, the vacuum state breaks translation symmetry. If we are interested in translation-symmetric vacua,  one may restrict either to Class 1 or to the Class 2 states with a constant, non-negative $\mathcal{Q_B}$ sector. That is, as if the vacuum state has a given uniform constant label (which is not a charge) in the background; see the comment below.

\item[(6)] By construction, none of the Class 2 states built upon the Class 2 vacuum state $ |\mathcal{Q_B}\rangle_+$ are ``charged'' in the usual sense: they are not eigenstates of $\boldsymbol{\mathcal{Q_B}}$. Nonetheless, they are  non-zero eigenstates of $(\boldsymbol{\mathcal{Q_B}})^2$ with eigenvalue  $(\mathcal{Q_B})^2$.

\item[(7)] Physical Hilbert spaces built upon any two arbitrary $\mathcal{Q_B}(\vec{x})$ and $\tilde{\mathcal{Q}}_{\cal B}(\vec{x})$ are in one-to-one correspondence. Moreover, any two states in any such two super-selection sectors of physical Hilbert spaces are orthogonal to each other.

\item [(8)] $ |\mathcal{Q_B}\rangle_+$ is the sum of two eigenstates of $\mathcal{Q_B}$ with opposite charge densities. Thus, in a loose sense, one may think of these states as having a vanishing ``total electric charge.'' For the special super-selection sector with constant ($x$-independent) $\mathcal{Q_B}$, there is an ``electric dipole'' associated with the vacuum state.
\item[(9)] Charge conjugation is not an operator well-defined over a Class 2 physical Hilbert space $\cHp$. However, it is physically expected that charge conjugation invariance should be retained for physical observables within QED, in which physical observables are of the form of gauge-invariant local operators. {An explicit demonstration of this fact follows directly from arguments and analysis similarly to those in section \ref{sec:derivation} that led to \eqref{n-point-Ca}.  }
\item[(10)] In our construction we dropped the soft photon part $\tilde{A}(\vec{k})$. Since the soft photon creation operator commute with both  $\boldsymbol{\epsilon}^\perp_a(k)$ and also $\boldsymbol{\mathcal{Q_B}}(\vec{x})$, we can include the soft photons in our analysis verbatim, in the same way that they are added to the Class 1 state, e.g. as in \cite{Strominger:2017zoo}.
\item[(11)] Since the theory is gauge invariant, it is expected that these super-selection sectors in a Class 2 Hilbert space do not mix with each other. This point should be more rigorously verified. We postpone this to future studies.
\item[(12)] The above discussions and analyses for Maxwell theory can be extended to non-Abelian gauge theories. One of the main points used in our construction is the existence of a $\mathbb{Z}_2$ symmetry \textbf{\sho}. In the non-Abelian theories one would expect that, as in the case of Maxwell theory/QED, charge conjugation should be a good choice. Moreover, in the non-Abelian case equations of motion in general and the constraint equation in particular, are non-linear equations. Nonetheless,  the sandwich constraint can be solve in terms of eigenvalues of $\boldsymbol{\mathcal{Q_B}}\equiv \vec{\nabla}\cdot\vec{\mathbf{E}}$ (and not directly in terms of gauge field components).  A detailed study of the non-Abelian theories should be carried out  in an independent analysis.

\item[(13)] In an Abelian gauge theory like Maxwell theory, one may add a genuine charge density, say $\boldsymbol{\mathcal{Q_G}}$, by simply  replacing $\boldsymbol{\mathcal{Q_B}}$ with $\boldsymbol{\mathcal{Q_B}}+\boldsymbol{\mathcal{Q_G}}$, explicitly, we need to solve the sandwich condition
$ \langle\psi|\vec{\nabla}\cdot\vec{\mathbf{E}}|\phi\rangle=\langle\psi|\boldsymbol{\mathcal{Q_G}}|\phi\rangle$ instead of \eqref{Maxwell-sandwich-2}. Since the constraint equation is linear in $\boldsymbol{\mathcal{Q_B}}$ and $\vec{\mathbf{E}}$, one can straightforwardly extend our construction of physical Hilbert spaces and solutions to sandwich equations  to include $\boldsymbol{\mathcal{Q_G}}$: The genuinely charged states may be constructed by the action of gauge invariant local operators on a generic vacuum state $|\mathcal{Q_B}\rangle_+ $.

\end{itemize}

{\section{Outlook}}\label{sec:conc}

In this short note, we revisited the old problem of the quantization of gauge field theories. We argued that there is a small, but potentially important point that has slipped the attention of physicists. The sandwich conditions were discussed in (very) early QED literature, see e.g. \cite{Gupta:1949rh, Bleuler:1950cy} or in the string textbook, see first page of chapter 4 in \cite{Polchinski:1998rq}. However, it seems to be overlooked or forgotten in the development of quantum gauge field theories. To the knowledge of the author, the reincarnation of sandwich quantization scheme in more recent literature happened with a different motivation and argument than the one presented here. It was first presented for the quantization of null strings, i.e. worldsheet theory of tensionless strings whose worldsheet is a 2d null surface \cite{Bagchi:2020fpr}. It was then noted that the same quantization scheme may be invoked for generic string worldsheet theory \cite{Bagchi:2024tyq} and also null $p$-brane theory \cite{Dutta:2024gkc}. In this note we put these previous examples in a broader framework: sandwich quantization scheme can/should be invoked for any field theory with local (gauge) symmetry.

{Our main motivation for revisiting the quantization of gauge theories, while in part has been better understanding of null strings or branes, has to do with the more fundamental and abstruse problem of \textit{the role of observers  in quantum theory}. This problem has been there since the introduction of quantum mechanics a century ago and has continued in quantum field theory and up to date. There has been a resurgence of the same conceptual issue in the context of quantum gravity and (quantum) cosmology. In dealing with usual gauge theories (like QED, QCD or standard model of particle physics) and as far as computation of $n$-point function of local gauge-invariant operators are concerned, sandwich quantization, as we discussed, seems to be more like an academic question does not add to the computed physical observables. However, in my opinion and understanding, it can have a more fundamental role in quantum gravity and cosmology. The resolution of questions regarding quantum gravity, in my viewpoint, lies in a fresh and better understanding of gauge theories (note that gravity is a gauge theory with diffeomorphisms as gauge symmetry). The sandwich quantization proposal should be viewed as an attempt in this direction. The idea is that each super-selection sector in the physical Hilbert space should be associated with different observers. In other words, the role of observer is carried or manifested in the Hilbert space she uses; see the discussions below on ``gauge equivalence principle''. }

We stressed the already known, but often overlooked, fact  that (e.g. see \cite{Bleuler:1950cy, Kowalski-Glikman:1992bnv, Polchinski:1998rq}) for quantization it is sufficient to impose the EoM for the gauge d.o.f, which appear as constraints, through sandwich conditions \eqref{Our-quatization-scheme}. We briefly discussed how the sandwich conditions may be solved, leading to a new set of super-selection sectors in the physical Hilbert space besides the textbook standard Class 1 states. The Class 1 physical Hilbert space, obtained by imposing gauge-fixing constraints through the right-action \eqref{RA-constraints}, is a direct extension of the classical constraints to the quantum level. The sandwich conditions and Class 2 solutions are the options arising through the quantization procedure, with no direct classical counterpart. The latter may be witnessed by the appearance of $\hbar$ in the right-hand side of \eqref{n-point-Ca-2}. At the quantum level, we have the option that, while $\bcC^i$ vanish when inserted into the $n$-point function of any generic local operators, the operator $\bcC^i$ need not be well-defined over the physical Hilbert space $\cHp$. In other words, requiring that $\bcC^i$ are well defined over $\cHp$ restricts us to the Class 1 Hilbert space, whereas relaxing this seemingly unnecessary requirement, opens up the possibility of super-selection sectors and the Class 2 physical Hilbert spaces. There are some steps are remaining to complete the sandwich quantization proposal and show its consistency and sufficiency.

\paragraph{Consistency and sufficiency of the proposal.} In this work, for illustration purposes we focused on the ``axial-type gauges''. Consistency of the proposal requires that it should work for any arbitrary gauge-fixing. For these cases one should work through introducing ghosts and check the BRST invariance, that the physical states satisfying the sandwich conditions are defined up to BRST equivalence classes (see e.g. \cite{Henneaux:1992} for discussion on BRST symmetry). That is,  we need to show that our each super-selection sector in the Class 2 physical Hilbert spaces is BRST invariant; in other words, one need to work through a ``sandwich version'' of the BRST symmetry and analysis.

\paragraph{Completion of the proposal.} We need to uncover the physical meaning of the new sectors in the physical Hilbert space, the super-selection sectors in the Class 2 part. First we note that, Class 2 states by definition, are orthogonal to the Class 1 states, and the gauge-invariant dynamics of the theory will never mix them with the Class 1 states. That is, one can consistently formulate the theory using only the Class 1 states and recover the standard QFT textbook results. There is, however, another option: physics may be formulated based on either of the super-selection sectors in the Class 2 Hilbert space. Based on the example explored in \cite{Bagchi:2024tyq} and the  Maxwell theory example discussed in the previous section, we propose the ``sandwich equivalence principle'':
\begin{center}
    \textit{Physical observables of gauge field theories should equivalently be described by the Class 1 or either of super-selection sectors in the Class 2 physical Hilbert space. }
\end{center}
The above is somewhat similar to what we have in Einstein's Equivalence Principle and the choice of different observers: each super-selection sector in the Class 2 physical Hilbert space provides states above a vacuum state associated with a particular physical observer. Of course, it remains to fully establish and formulate this equivalence principle. {Conversely and recalling that Einstein's general relativity is the gauge theory with diffeomorphism as the gauge symmetry, the gauge equivalence principle stated above may be viewed as a proposal for taking Einstein's equivalence principle to quantum level. Noting that fixing diffeomorphisms is physically corresponding to choosing an observer, the ``gauge equivalence principle''  highlights the role of observer in quantum gravity and provides a framework to formulate it. }

A lot more should still be done to develop the formulation and establish consistency and sufficiency of the sandwich equivalence principle stated above, and to verify what new insights on quantization our new scheme brings and whether the gained insights can be useful in addressing interesting physics questions. Two examples could be of particular interest: quantization of worldline theory of a particle, as a $0+1$ dimensional theory which enjoys the worldline reprarametrization invariance as gauge symmetry. This is the simplest setup that allows for an explicit and detailed analysis. This exercise is currently under study \cite{Worldline-theory}. The other is the example of general relativity with spacetime diffeomorphism as gauge symmetry. This analysis can provide a new venue for quantization of general relativity, beyond the Wheeler-DeWitt framework \cite{DeWitt:1967yk} and may lead to a generally covariant arrow of time as a quantum feature \cite{AoT-Shahin}.

\section*{Acknowledgement}
I would like to thank Arjun Bagchi, Aritra Banerjee, Sergio Cecotti and especially Anton Pribytok, Ida Rasulian and  Hossein Yavartanoo, for many discussions on the sandwich quantization scheme for strings, null branes and point particle. I would like to thank Glenn Barnich,  Marc Henneaux, Farhang Loran, Michael Peskin and Marco Serone for comments on the manuscript.  The author acknowledges support from INSF Research Chair No. 4045163.

\bibliographystyle{fullsort.bst}
\bibliography{reference}


	\end{document}